\documentclass[epsfig,12pt]{article}
\usepackage{epsfig}
\usepackage{graphicx}

\usepackage{latexsym}
\usepackage{amsmath}
\usepackage{amssymb}
\usepackage{relsize}
\usepackage{geometry}
\usepackage{hyperref}
\usepackage [T2A] { fontenc }
\usepackage[utf8]{inputenc}
\usepackage [english] {babel}
\usepackage{amsmath,amssymb}

\usepackage[numbers]{natbib}

\def \sec{\begin{section}}
\def \esec{\end{section}}

\def \la {\lambda}
\def \La {\Lambda}

\def \pr {\partial}
\def \ra {\rightarrow}
\def \qs {\qquad}

\def \l {\left(}
\def \r {\right)}
\def\beq{\begin{equation}}
\def\eeq{\end{equation}}
\def\beqn{\begin{eqnarray}}
\def\eeqn{\end{eqnarray}}

\newcommand{\pt}{\partial}

\newcommand{\gsim}{\lower.7ex\hbox{$
\;\stackrel{\textstyle>}{\sim}\;$}}
\newcommand{\lsim}{\lower.7ex\hbox{$
\;\stackrel{\textstyle<}{\sim}\;$}}

\def\slashed#1{\setbox0=\hbox{$#1$}             
   \dimen0=\wd0                                 
   \setbox1=\hbox{/} \dimen1=\wd1               
   \ifdim\dimen0>\dimen1                        
      \rlap{\hbox to \dimen0{\hfil/\hfil}}      
      #1                                        
   \else                                        
      \rlap{\hbox to \dimen1{\hfil$#1$\hfil}}   
      /                                         
   \fi}                                        %

\begin{document}


\begin{titlepage}

\begin{flushright}
ITEP-TH- 12/13\\
\end{flushright}

\vspace{1cm}

\begin{center}
{  \Large \bf The $CP(N-1)$ model on a Disc and Decay of a Non-Abelian String }
\end{center}
\vspace{1mm}

\begin{center}

 {\large
 \bf   A.~Gorsky$\,^{a,b}$ and A.~ Milekhin$\,^{a,b}$}

\vspace{3mm}

$^a$
{\it Institute of Theoretical and Experimental Physics,
Moscow 117218, Russia}\\[1mm]
$^b$
{\it Moscow Institute of Physics and Technology,
Dolgoprudny 141700, Russia}
\end{center}

\centerline{\small\tt gorsky@itep.ru }
\centerline{\small\tt milekhin@itep.ru }

\vspace{1cm}

\begin{center}
{\large\bf Abstract}
\end{center}

We consider the role of quantum effects in the non-perturbative  decay
of non-abelian string with orientational moduli in non-supersymmetric D=4 gauge  theory.
To this aim the effective action in the $CP(N-1)$ model on a disc at large N has been
calculated. It exhibits  phase transition at some radius, the "wrong sign" Luscher term
and  large boundary boojum-like negative contribution. The effect of $\theta$ - term and the possibility
of the spontaneous creation of the non-abelian string are briefly discussed.

\end{titlepage}


\section{Introduction}

 The non-abelian strings \cite{tong,konishi,sy} during the last decade become the
prominent element in the landscape
of the world of non-perturbative extended objects in supersymmetric and
non-supersymmetric \cite{gsy}
gauge theories with the different matter content. When tension of the nonabelian
string tends to infinity it is known under the name of the surface operator providing the
proper boundary conditions on some hyperplanes.
 Unlike the  Abrikosov string(abelian string)
which possesses only 2 translation moduli, non-abelian strings possess additional
 internal orientational degrees of freedom  corresponding to the nontrivial
embedding of the string in the peculiar global group.
In the simplest non-supersymmetric model \cite{gsy} the  low energy dynamics
of  orientational moduli is described by the $CP(N-1)$ sigma model
(see \cite{Shifman:2009zz} for a review).

In this note we shall focus on the peculiar issue concerning the decay of nonabelian string
with orientational moduli. The decay of the abelian strings has been considered in the
cosmological context  \cite{vilenkin1,vilenkin2} and in the
specific gauge theories \cite{sy2002}.
The internal moduli makes the decay process more involved. This process has been
briefly discussed in \cite{gorvol2010} in the semiclassical setting. Here
we shall take into account the quantum effects in the worldsheet theory
modifying some aspects of the decay considerably.
The worldsheet $CP(N-1)$ sigma model can be studied via 1/N expansion
and possesses, for example, dynamical mass generation, asymptotic freedom, confinement,
non-trivial $\theta$-angle dependence(\cite{adda}, \cite{witten}).

Recently it was shown that the theory manifests
a phase transition upon  some deformation \cite{gorsky2006}or  at the interval \cite{milekhin2012}
Moreover the theory defined at the interval  shows
Casimir-type scaling in the vacuum energy despite
of the mass gap \cite{milekhin2012}. The sign of the Luscher
term is unusual but coincides with the one in some other models
with the mass gap \cite{zhit}. Another interesting point to be mentioned
is that there are  negative  edge contributions to the energy of the finite
non-abelian string \cite{milekhin2012}.

The decay process  from the worldsheet viewpoint is similar to the false vacuum decay in two dimensions. However
there are interesting features in this case. The Euclidean bounce in 1+1 corresponds to the $CP(N-1)$ sigma model
on the disc and it turns out that the model on the disc undergoes the phase transition similar to
one found in \cite{milekhin2012} in the strip geometry.
Therefore the decay rate is sensitive to the radius of the bounce which is fixed
by the parameters of the problem. Moreover in addition to the false vacuum decay  situation
we shall find in a large N approximation the
Luscher-type term  affecting the probability rate.

In the $CP(N-1)$ sigma model the bulk
$\theta$ term  generates the  topological term in the worldsheet theory \cite{gsy}. One could also
consider axions both in the bulk and worldsheet theory. It was found in \cite{axion}
that the worldsheet axion amounts to the deconfinement in the worldsheet theory.
We shall discuss the effects of the $\theta$-term and axion on the string decay rate.
.

The paper is organized as follows.
In Section 2 calculate the effective action in $CP(N-1)$ sigma model on the disc and show some
unusual properties. In particular we demonstrate the phase transition in the $CP(N-1)$ sigma model  
at some disc radius. In Section 3 we recall the simplest field theory model supporting non-abelian string.
In Section 4  we derive the  rate of the non-abelian string decay. We will also briefly discuss the decay induced by the kink-antikink meson living on the string
worldsheet. In Section 5 we discuss the impact of the $\theta$-term and the axion
on the decay process.  The final remarks can be found in the Conclusion.

\section{CP(N-1) model on the disc}
In this Section we shall perform some preliminary work
for the decay probability rate calculation
and derive the effective action for the $CP(N-1)$ model
on the disc geometry with Dirichlet boundary conditions.
We assume the sharp boundary of  the disc  which is
equivalent to the thin-wall approximation in the false vacuum decay.

Recall that $CP(N-1)$ sigma model is described by Lagrangian
\begin{equation}
\mathcal{L}=\cfrac{N}{g^2}(\pr_\mu-iA_\mu)n_i(\pr^\mu+iA^\mu)n^{*i}-\la(n_i^* n^i-1)
\end{equation}
where $i=1,...,N$  and $\la$ and $A_\mu$ are Lagrange multipliers. $\la$
impose the constraint $n_i^* n^i=1$, $A_\mu$ is just a dummy field that
could be eliminated by equation of motion $A_\mu=i n_i^* \pr_\mu n^i$
but makes U(1) invariance obvious.
In what  follows the
calculations are made in Euclidean space. As was mentioned above,
we  impose Dirichlet boundary conditions:
\begin{equation}
n^i=0, i=2,..,N; n^1=1,  \qs \sqrt{x_0^2+x_1^2}=R
\end{equation}

General strategy for the large N expansion is to integrate over $n_i$ and then
use the saddle-point approximation which is justified
by the large N. In the leading order
we could neglect the difference between $N$ and $N-1$.
Also we will assume that in the saddle-point $A_\mu=0$ and $\la=const$.
After the appropriate rescaling:
\begin{equation}
S_{eff}=N \log det(-\pr^2+m^2)-\int d^2x \la
\end{equation}
where $m^2=\cfrac{\la g^2}{N}$.
In $\mathbb{R}^2$ the determinant can be calculated easily
and the saddle-point equation $\cfrac{\pr S_{eff}}{\pr m}=0$
yields via the dimensional transmutation  the IR scale
\begin{equation}
m=\La_{uv} \exp{\cfrac{-2 \pi}{g^2}}
\end{equation}

Calculating the determinant in the disc geometry is  more
complicated since explicit expression for eigenvalues involves roots of
Bessel functions. Instead, we will use Gelfand-Yaglom method,
which could be generalized for a multi-dimensional case.
We can separate:
\begin{equation}
\label{eq:det}
\log{det(-\pr^2+m^2)} = \log{det(-\pr^2)} - \log{\cfrac{det(-\pr^2)}{det(-\pr^2+m^2)}}
\end{equation}
$\log{det(-\pr^2)}$ - is irrelevant constant.

According to \cite{dunne}, in two dimensions with Dirichlet boundary conditions:

\begin{eqnarray}
\label{eq:maineq}
\log{\cfrac{det(M+m^2)}{det(M^{free}+m^2)}} = \log{\cfrac{\psi_0(R)}{\psi_0^{free}(R)}} + 2 \sum_{l=1}^{\infty} \l
\log{\cfrac{\psi_l(R)}{\psi_0^{free}(R)}} -
\cfrac{1}{2l} \int_0^R r V(r) dr \r \\ \nonumber
+ \int_0^R r V(r) \l \log{\cfrac{\mu r}{2}} + \gamma  \r \
\end{eqnarray}
where M - is Schrodinger-like operator
\begin{equation}
M = -\pr^2  + V(r)
\end{equation}
\begin{equation}
M^{free} = -\pr^2
\end{equation}
and
\begin{equation}
(-\pr_r^2 + \cfrac{l^2}{r^2} + V(r) + m^2) \psi_l(r) = 0, \qs \psi_l(r) \approx r^{l}, as \qs r \ra 0.
\end{equation}
$\psi_l^{free}$ is defined similarly with the $V(r)$ omitted.
$\gamma \approx 0,577...$ - Euler-Mascheroni constant and $\mu$-is
renormalization scale - that is (\ref{eq:maineq}) is free of
divergences. $\gamma+\log{1/2}$ could be eliminated through the
choice of $\mu$. Actually, (\ref{eq:maineq}) is derived using
zeta-function regularization for slightly modified determinant
$det(\cfrac{...}{\mu^2})$ which is equivalent to renormalization at scale $\mu$.
That is  we should write
$\cfrac{N \pi (mR)^2}{g_\mu^2}$ instead of $\cfrac{N \pi (mR)^2}{g^2}$

In our case $V(r)=-m^2$, hence
\begin{equation}
\cfrac{\psi_l(r)}{\psi_l^{free}(r)} = \cfrac{(mr)^l}{I_l(mr) l! 2^l}
\end{equation}
where $I_l(r)$ - is Infeld function(modified Bessel function of second kind).

Therefore we can represent
 (\ref{eq:det}) in the following form:
\begin{equation}
\label{eq:maineq2}
\log(I_0(mR))+2 \sum_{l=1}^{\infty} \l \log{\cfrac{I_l(mR) l! 2^l}{(mR)^l}} -
\cfrac{(mR)^2}{4l} \r + \cfrac{(mR)^2}{2} \log(\mu R) - \cfrac{(mR)^2}{4}
\end{equation}
and consider consider two limiting cases: $mR>>1$ and $mR<<1$.
The later is  more simple, because Taylor expansion of $I_l$
in the vicinity of 0 is valid for all $l$ if $mR<<1$, while
WKB-asymptotics $I_l \approx e^x$ is valid only for $mR>>l^2$ which,
of course, is not true for sufficiently large l.

At small  $mR=x$. the effective action reads as
\begin{equation}
S_{eff}= N \l -\cfrac{x^2}{2}-x^4 \l \cfrac{\pi^2-9}{96}+\cfrac{1}{64} \r+\cfrac{x^2}{2} \log{\mu R} +
O(x^6) \r -\cfrac{N \pi x^2}{g_\mu^2}
\end{equation}
and
\begin{equation}
\label{eq:dimtrans}
\La=\mu \exp{\cfrac{- 2\pi}{g_\mu^2}}
\end{equation}
is dynamically generated  scale.
The saddle point equation
with account of  (\ref{eq:dimtrans}) yields
\begin{equation}
2 x^2  \l \cfrac{\pi^2-9}{96}+\cfrac{1}{64} \r = -\cfrac{1}{2}+\cfrac{1}{2}\log{\La R}
\end{equation}
This equation does not have solutions for the sufficiently
small R which  indicates  the phase transition at $R \approx 1/\La$.
Actually it means that $n^1$ receives non-zero VEV.
The situation
is very similar to one discussed in \cite{milekhin2012}:
Dirichlet boundary conditions broke global $SU(N)$ to $SU(N-1)$.

Now let us investigate the case $x>>1$.
The numerical calculations give
\begin{equation}
S_{eff}=N \l -\cfrac{x^2}{2} \log{x} + A x^2 - B x +
C \log{x}+\cfrac{x^2}{2} \log{\mu R}-\cfrac{x^2}{4}\r -\cfrac{N \pi x^2}{g_\mu^2}
\end{equation}
with $A\approx 1/2, B\approx 1.6, C\approx 1$
Saddle point equation
\begin{equation}
\cfrac{\pr S_{eff}}{\pr x^2}=0=-\cfrac{1}{2} \log{x}-\cfrac{1}{2}+A-\cfrac{B}{2x}+
\cfrac{C}{2x^2}+\cfrac{1}{2} \log{\La R}
\end{equation}
yields
\begin{equation}
x=m R = \La R -B  + \cfrac{C-B^2/2}{\La R} + O(1/R^2)
\end{equation}
and
\begin{equation}
S_{eff}=N \l \cfrac{\La^2 R^2}{4} - B \La R + C \log{\La R} + O(1) \r
\label{action}
\end{equation}
The area term $\La^2 R^2/4$ reproduces the result for the
flat $\mathbb{R}^2$ case (\cite{novikov}).

The perimeter term reflects the contribution from the boundary excitation while the log term
can be thought of as the contribution from the Luscher term. Let us emphasize that
the signs of the different contributions are similar to the ones obtained in the model at the interval.
The signs are unusual however and deserve for the comments. First,
the Luscher logarithmic term has the sign which is opposite to the standard one
but coincides with the results obtained in the other theories with the
mass gap \cite{zhit}. Let us emphasize that the previous discussion
on this term in $CP(N-1)$ model \cite{luscher} seems to be  incomplete. It was based on the argument
that the mass gap results into the exponential suppression of the Luscher
term resulted from the orientational moduli however it turns out that this argument does not work.

Another surprising point concerns the sign of the boundary contribution
which corresponds to the localized negative energy of order O(N). At the first
glance this looks extremely suspicious however it turns out that such boojums
with the negative energy are known in some solid state examples and field theory
models. In the most close situation the boojums occur at the intersection
of the strings and domain walls \cite{bo1,bo2,bo3} where the Dirichlet boundary
conditions are imposed by domain walls.  The detailed analysis in \cite{bo1,bo2,bo3}
demonstrated that the negative edge energy is consistent with the equation
of motion and even with supersymmetry.

Let us emphasize that the boojum contribution in \cite{bo1,bo2,bo3} was evaluated
at weak coupling which was provided by the large twisted mass on the string worldsheet.
The semiclassical consideration yields in $SU(2)$ case the boundary energy equals to the half of the
kink mass with the opposite sign. In our case we consider the pure quantum problem however for
$CP(1)$ case we reproduce at strong coupling the same value of the boundary boojum energy. At large N
the edge contribution is proportional to N. It would be very interesting to get the better
physical explanation of the boojum negative energy.

In the  Section 5 we shall use the effective
action in the disc geometry to get the  rate of the string decay.

\section{Non-abelian strings}

Here we review the simplest model  with the gauge group SU($N)\times$U(1).
which can be used
to analyze non-abelian strings \cite{gsy}.
The model contains $N$ scalar fields charged with respect to
$U(1)$ which form $N$ fundamental representations of SU($N$).
It is convenient to write these fields as
$N\times N$ matrix $\Phi =\{\varphi^{kA}\}$
where $k$ is the SU($N$) gauge index while $A$ is the flavor
index,
\beq
\Phi =\left(
\begin{array}{cccc}
\varphi^{11} & \varphi^{12}& ... & \varphi^{1N}\\[2mm]
\varphi^{21} & \varphi^{22}& ... & \varphi^{2N}\\[2mm]
....&...&...&...\\[2mm]
\varphi^{N1} & \varphi^{N2}& ... & \varphi^{NN}
\end{array}
\right)\,.
\label{phima}
\eeq
The action of the model has the form\,
\beqn
S &=& \int {\rm d}^4x\left\{\frac1{4g_2^2}
\left(F^{a}_{\mu\nu}\right)^{2}
+ \frac1{4g_1^2}\left(F_{\mu\nu}\right)^{2}
 \right.
 \nonumber\\[3mm]
&+&
 {\rm Tr}\, (\nabla_\mu \Phi)^\dagger \,(\nabla^\mu \Phi )
+\frac{g^2_2}{2}\left[{\rm Tr}\,
\left(\Phi^\dagger T^a \Phi\right)\right]^2
 +
 \frac{g^2_1}{8}\left[ {\rm Tr}\,
\left( \Phi^\dagger \Phi \right)- N\xi \right]^2
 \nonumber\\[3mm]
 &+&\left.
 \frac{i\,\theta}{32\,\pi^2} \, F_{\mu\nu}^a \tilde{F}^{a\,\mu\nu}
 \right\}\,,
\label{redqed}
\eeqn
where $T^a$ stands for the generator of the gauge SU($N$),
\beq
\nabla_\mu \, \Phi \equiv  \left( \partial_\mu -\frac{i}{\sqrt{ 2N}}\; A_{\mu}
-i A^{a}_{\mu}\, T^a\right)\Phi\, ,
\label{dcde}
\eeq
and $\theta$ is the vacuum angle.  The last
term in the second line
forces $\Phi$ to develop a vacuum expectation value (VEV) while the
last but one term
forces the VEV to be diagonal,
\beq
\Phi_{\rm vac} = \sqrt\xi\,{\rm diag}\, \{1,1,...,1\}\,.
\label{diagphi}
\eeq

We assume that the parameter $\xi$ to be large,
\beq
\sqrt{\xi}\gg \Lambda_4,
\label{weakcoupling}
\eeq
where $\Lambda_4$ is the scale of the four-dimensional theory (\ref{redqed}).
That is we are in the weak coupling regime as both couplings $g^2_1$ and
$g^2_2$
are frozen at a large scale.

The  vacuum field (\ref{diagphi}) results in  the spontaneous
breaking of both gauge and flavor SU($N$)'s.
A diagonal global SU($N$) survives
\beq
{\rm U}(N)_{\rm gauge}\times {\rm SU}(N)_{\rm flavor}
\to {\rm SU}(N)_{\rm diag}\,.
\eeq
yielding color-flavor locking  in the vacuum.

The nontrivial topology providing the stability is based on
\beq
\pi_1 \left({\rm SU}(N)\times {\rm U}(1)/ Z_N
\right)\neq 0\,.
\eeq
and one can wind   one element of $\Phi_{\rm vac}$,
\beq
\Phi_{\rm string} = \sqrt{\xi}\,{\rm diag} ( 1,1, ... ,e^{i\alpha (x) })\,,
\quad x\to\infty \,.
\label{ansa}
\eeq
The strings with  this boundary condition can be called elementary $Z_N$ strings;
their tension is $1/N$-th of that of the ANO string.
The ANO string can be viewed as a bound state of
$N$ $Z_N$ strings.
The $Z_N$ string solution
can be written as
follows :
\beqn
\Phi &=&
\left(
\begin{array}{cccc}
\phi(r) & 0& ... & 0\\[2mm]
....&...&...&...\\[2mm]
0& ... & \phi(r)&  0\\[2mm]
0 & 0& ... & e^{i\alpha}\phi_{N}(r)
\end{array}
\right) ,
\nonumber\\[5mm]
A^{{\rm SU}(N)}_i &=&
\frac1N\left(
\begin{array}{cccc}
1 & ... & 0 & 0\\[2mm]
....&...&...&...\\[2mm]
0&  ... & 1 & 0\\[2mm]
0 & 0& ... & -(N-1)
\end{array}
\right)\, \left( \pt_i \alpha \right) \left[ -1+f_{NA}(r)\right] ,
\nonumber\\[5mm]
A^{{\rm U}(1)}_i &=& \frac{1}{N}\,
\left( \pt_i \alpha \right)\left[1-f(r)\right] ,\qquad A^{{\rm U}(1)}_0=
A^{{\rm SU}(N)}_0 =0\,,
\label{znstr}
\eeqn
where $i=1,2$ labels coordinates in the plane orthogonal to the string
axis and $r$ and $\alpha$ are the polar coordinates in this plane. The profile functions
satisfy the following
boundary conditions:
\beqn
&& \phi_{N}(0)=0,
\nonumber\\[2mm]
&& f_{NA}(0)=1,\;\;\;f(0)=1\,,
\label{bc0}
\eeqn
at $r=0$, and
\beqn
&& \phi_{N}(\infty)=\sqrt{\xi},\;\;\;\phi(\infty)=\sqrt{\xi}\,,
\nonumber\\[2mm]
&& f_{NA}(\infty)=0,\;\;\;\; \; f(\infty) = 0
\label{bcinfty}
\eeqn
at $r=\infty$.

The tension of this elementary string is
\beq
T_1=2\pi\,\xi\, .
\label{ten}
\eeq
while the tension of
the ANO string is
\beq
T_{\rm ANO}=2\pi\,N\,\xi
\label{tenANO}
\eeq
which confirms its composite nature.

To obtain the non-Abelian string solution from the $Z_N$ string
(\ref{znstr}) we apply the diagonal color-flavor rotation  preserving
the vacuum (\ref{diagphi}).
In singular gauge we have
\beqn
\Phi &=&
U\left(
\begin{array}{cccc}
\phi(r) & 0& ... & 0\\[2mm]
....&...&...&...\\[2mm]
0& ... & \phi(r)&  0\\[2mm]
0 & 0& ... & \phi_{N}(r)
\end{array}
\right)U^{-1}\, ,
\nonumber\\[5mm]
A^{{\rm SU}(N)}_i &=&
\frac{1}{N} \,U\left(
\begin{array}{cccc}
1 & ... & 0 & 0\\[2mm]
....&...&...&...\\[2mm]
0&  ... & 1 & 0\\[2mm]
0 & 0& ... & -(N-1)
\end{array}
\right)U^{-1}\, \left( \pt_i \alpha\right)  f_{NA}(r)\, ,
\nonumber\\[5mm]
A^{{\rm U}(1)}_i &=& -\frac{1}{N}\,
\left( \pt_i \alpha\right)   f(r)\, , \qquad A^{{\rm U}(1)}_0=
A^{{\rm SU}(N)}_0=0\,,
\label{nastr}
\eeqn
where $U$ is a matrix $\in {\rm SU}(N)$. This matrix parameterizes
orientational zero modes of the string associated with flux embedding
into  SU($N$).

Let us discuss the worldsheet description of the nonabelian string.
It is important that there are two independent contribution
from "`space"' and "`internal"' terms.
To obtain the   kinetic term in the "`internal"' action we follow the standard
logic in the derivation of the low-energy
action in the moduli approximation. That is we substitute our solution, which
depends
on the moduli $ n^l$, in the action , assuming  that
the fields acquire a dependence on the coordinates $x_k$ via $n^l(x_k)$.
Then we arrive at the $CP(N-1)$  sigma model ,
\beq
S^{(1+1)}_{CP(N-1)}= 2 f\,   \int d t\, dz \,  \left\{(\pt_{k}\, n^*
\pt_{k}\, n) + (n^*\pt_{k}\, n)^2\right\}\,,
\label{cp}
\eeq
where the coupling constant $f$ is given by a normalizing integral
defined in terms of the string profile functions which yields

\beq
f= \frac{2\pi}{g_2^2}\,.
\label{betag}
\eeq
that is two-dimensional coupling constant is determined by the
four-dimensional non-Abelian coupling.

The sigma model (\ref{cp}) is asymptotically free hence at large
distances it gets into the strong coupling regime.  The  running
coupling constant  as a function of the energy scale $E$ at one loop is given
by
\beq
4\pi f = N\ln {\left(\frac{E}{\Lambda_{CP(N-1)}}\right)}
+\cdots,
\label{sigmacoup}
\eeq
where $\Lambda_{CP(N-1)}$ is the dynamical scale of the $CP(N-1)$
model.
The UV cut-off of the sigma model at hand
is determined by  $g_2\sqrt{\xi}$,
hence,
\beq
\Lambda^N_{CP(N-1)} = g_2^N\, \xi^{N/2} \,\, e^{-\frac{8\pi^2}{g^2_2}} .
\label{lambdasig}
\eeq
In the bulk theory, due to the VEV's of
the scalar fields, the coupling constant is frozen at
$g_2\sqrt{\xi}$. There are no logarithms in the bulk theory
below this scale and the logarithms of the
world-sheet  theory take over.

\section{Decay of the string}
\subsection{Decay  rate}
Consider now the decay  rate. In the simple ANO -like abelian string the probability
can be estimated
using the total effective action for the classical Euclidean bounce(hole)
\beq
S_{eff}^{U(1)}= T_0\pi R^2 - M_0 2\pi R
\eeq
where the origin of two terms is evident. The logarithmic Luscher
term in the abelian case can be neglected with. The extremization with respect to the radius of the
hole yields the value of the effective action evaluated at the critical bounce. Taking into
account the exact answer for the preexponential factor in 1+1 dimensional false vacuum decay
calculated in \cite{kissel,vol} the probability rate reads as
\beq
w=\frac{T_0}{2\pi} \exp(-\frac{M_0^2}{T_0})
\label{w}
\eeq

There are a few new issues for  the non-abelian string. The total
effective action involves now two terms
\beq
S_{eff} = S_{eff}^{U(1)} + S_{eff}^{CP(N-1)}
\eeq
As we have shown in Section 2 there are two phases in the  $CP(N-1)$ model and
the $S_{eff}^{CP(N-1)}$ depends on the solution $R_{crit}$
to the extremization equation. If the $R_{crit}$ is large  one could use
the asymptotic behavior (\ref {action}) and the total action reads as
\beq
S_{eff}= T_{eff} \pi R^2 - M_{eff} 2\pi R + NC \log \Lambda R
\eeq
where we neglect small abelian contribution to the Luscher term and
\beq
T_{eff} = T_0 + N\Lambda^2/4, \qquad M_{eff} = M_0 - NB\Lambda.
\eeq
The validity of this approximation can be justified with the proper
choice of the parameters from the abelian sector. In this regime
the abelian component of the string dominates which however
needs for the very large values of the abelian tension and mass $T_0,M_0$. The decay
rate has the form (\ref{w}) corrected by the
additional preexponential factor due to the Luscher term.

It is interesting to discuss the opposite limit when the
orientational moduli dominates that is question
if there are stable Euclidean configurations with the radius near the phase transition .
The saddle point
solution for the non-abelian part of the action with respect to the radius can be analyzed exactly because the mass gap is also obtained
as the extremum of the action  with respect to the mass. Indeed the action has the form($C_1,C_2$ are some constants)
\beq
S=\log{det(-\pr^2+m^2)}-C_1 mL/g^2 = \log{det(-\pr^2+m^2)}_{reg} + C_2 m R \log{\La R}
\eeq
where
$\log{det(-\pr^2+m^2)}_{reg}$ - is well-defined function of $mR$ only. $m(R)$ is the solution to the equation
\beq
\cfrac{\pr S(m,R)}{\pr m}=0
\eeq
Critical radius is the solution to
\beq
\cfrac{d S(m(R),R)}{d R}=0
\eeq
Now it is not difficult to see that this equation reduces to $m=0$, that is, the critical radius coincides
with the phase transition radius.
When the abelian part of the action is added only the numerical analysis is available. It demonstrates
that by the choice of the abelian parameters the critical radius could move in both directions
from the transition point. However if it is less then $R_{tran}$ the analysis becomes inconsistent.

One could also imagine more involved situation when only the non-abelian
flux terminates at the boundary of the disc on the string worldsheet
in the Euclidean space
while the abelian flux remains intact.
To some extend such process would mean that string non-perturbatively drops off
its nonabelian hair and becomes the conventional abelian string.
This means that the  boundary excitations with the negative energy  are
created and propagate along the string in the opposite directions separating two
regions.

\subsection{Spontaneous creation of the string}

One could also comment on the opposite problem. Namely let us assume that we are
in the vacuum state in the 4D theory which admits the non-abelian string and question
if the non-abelian string can be created spontaneously. The question
is reasonable since there are competing area and perimeter terms  in the  $CP(N-1)$ model
in the disc geometry.
It looks a bit unusual since contrary to the standard case the boundary term
in the energy of the long string is negative. In the standard case both area
and perimeter terms are positive and the saddle point is absent.

If  we consider the effective action of the  $CP(N-1)$ model
on the disc discussed above which involves area, perimeter and Luscher terms
the saddle point equation for the radius of the disc has to be found. In the absence
of the abelian component the critical radius coincides with the phase transition point
and the claim on the existence of the consistent saddle point  can not be made. Hence
the abelian component has to be taken into account. It is necessary to take $M_0$
small enough to keep the effective perimeter term negative. The numerical analysis demonstrates
that the saddle point does exist in some region of the parameter space.

Let us question on the related problem. Is there the stable finite length
non-abelian string? If it exists it could be considered as the part of the spectrum.
To answer this question consider the extremization equation for the
energy which involves the abelian terms and the non-abelian contribution
found in \cite{milekhin2012}.
\beq
E = T_0 L + 2 M_0 + E_{eff}^{CP(N-1)}(L)
\eeq
It turns out that the solution to the extremization equation does not
exist hence there are no 
short magnetic strings in the spectrum.

\subsection{Induced decay}

Another issue concerns the induced decay of the non-abelian string. This process is
analogous to the induced false vacuum decay  in two dimensions
\cite{induced,gorvol2005}. The decay can be induced by the particle excitation above the vacuum state.
The same process can be interpreted as the non-perturbative decay of the
particle in the false vacuum. The physical reason for the enhanced
decay is as follows. The external particle usually has zero mode on the kink therefore
the initial particle gets localized as zero mode at the boundary of the bounce
and the tunneling occurs not at zero but at energy equals the particle mass. The
bounce itself gets deformed into a fish-like configuration with two cusps.

In our case the decay can be induced by the kink-antikink meson.
The classical action for the deformed bounce solution reads as \cite{induced}
\beq
S_{ind}= \frac{2M^2}{T}arcsin\frac{m}{2M} - \frac{mM}{T}\sqrt{1- \frac{m^2}{4M^2}}
\label{induced}
\eeq
where $m$ is of order of the meson mass and M is the mass of the boundary state.
The preexponential factor has been calculated as well \cite{gorvol2005}.
In our case the induced process  could be interpreted as the decay of the  meson into
constituents. The kink and antikink get localized at the hole boundary. The kink masses
are of order $\Lambda$ hence we can certainly claim that the tunneling happens
at the energy of order $\Lambda$  which has to be substituted  as the mass of the
external particle in (\ref{induced}). However more detailed analysis is required
to get the induced decay rate.

\section{Axion and $\theta$-term}

In this Section we shall discuss how the $\theta$ term and axion affect the
decay rate. First, let us recall the results of \cite{axion}.
Two key properties were found; the 2d axion results in the deconfinement phase transition on the string
worldsheet while the 4d axion does not.
On the other hand  the four-dimensional $\theta$-term penetrates into
the worldsheet theory
\beq
\theta_{D=4}= \theta_{D=2}
\eeq
The mechanism of deconfinement is quite transparent.
The worldsheet  Lagrangian
involves  the photon-axion mixing. Upon taking into account this mixing
the propagating particle becomes massive and there is no linear confinement  between
kink and antikink. Therefore if the worldsheet axion is present the kink is liberated and
becomes the particle in the physical spectrum of the worldsheet theory. The deconfinement
phase transition has happened at $\theta =\pi$ as well due to the degeneration of the
ground state of the worldsheet theory. Let us also mention that kinks became the dyonic
particles if the $\theta$ term is included.

Turn now to the decay of the non-abelian string. The $\theta$ -term in Minkowski space provides
the constant
electric field along the string in the gauged formulation of the $CP(N-1)$ model similar to
\cite{coleman} hence
one could ask how it is screened at the ends of the string at the hole boundary. The  naive answer
would be  that
dyonic kinks are created at the ends of the string and screen the effective electric field.
However as we have seen above the boundary state has negative energy and does not have
naive kink interpretation. Nevertheless this boundary energy is proportional
to the non-perturbative scale therefore it becomes  $\theta$-dependent when the $\theta$-term
is switched on. This $\theta$-dependent boundary state should screen the electric
field along the string. We plan to discuss this subtle issue elsewhere.

In the $\theta$ enriched theory there is the non-vanishing vacuum  density of the
topological charge
\beq
<F>=\frac{dlogZ}{d\theta}\propto \Lambda^2 \sin(\frac{\theta}{N})
\eeq
Therefore one could ask what happens with the topological charge stored in the disc
which is removed in
the Euclidean bounce solution $Q_{top}=<F>\pi R^2 $. Upon the rotation
to the Euclidean space the electric field gets transformed into the effective magnetic field
transverse to the string worldsheet. There are two $\theta$-dependent boundary terms:
the screening charge which provides the Wilson loop along the boundary and
the $\theta$-dependent contribution to the boundary energy. We expect that
these terms are collected together to satisfy the Stokes
theorem
\beq
P \exp \oint _{bound}A = \exp (<F>R^2)
\eeq
Note that at large $N$ we can expand $sin(\frac{\theta}{N})$ dependence and
keep the linear term. This fits with the linear $\theta$-dependence of the
effective electric field. However to make these arguments precise it is necessary
to get the exact $\theta$ -dependence of the effective action.

Hence the total effect of the $\theta$ term is twofold. First,
the vacuum energy density is $\theta$-dependent therefore the decay rate is modified. Secondly,
there are dyonic-like boundary states  and there is nontrivial Wilson
loop of the auxiliary gauge field along the hole  boundary.
If the axion field localized at the string worldsheet is included from the very beginning
then one has to take into account the axion-photon mixing when
calculating the  vacuum energy. The decay is also accomplished by the strong axion
emission discussed in \cite{axion}.

Note in this context  that the $CP(N-1)$ model in Euclidean space was
used for description of the QHE droplets  at  strong magnetic field  \cite{bur}.
In that description the $\theta$ term provides the filling fraction in QHE $\frac{\theta}{2\pi}=\nu$ 
or equivalently the Hall conductivity.  The phase transition at $\theta=\pi$ is assumed to be related
with the transition betweem the platoes. The most interesting issue concerns the possible relation
between the boojum-like energy with the edge state in QHE and the possible phase transition at
 some radius of the droplet we have found. We hope to discuss these issues elsewhere.

\section{Conclusion}

In this note we have considered the quantum features  of non-abelian string decay in non-supersymmetric
gauge theory. To this aim the effective action of the $CP(N-1)$ model on  Euclidean disc
has been calculated and the phase transition similar to  one discovered  in \cite{milekhin2012} has been
found. The boundary boojum energy has been identified and the decay  rate
has been evaluated.  We also comment on  the
opposite process of  spontaneous creation of a non-abelian string.

The calculation of the action in the disc geometry shows a few general phenomena.
First, the Luscher logarithmic term has the "wrong" sign - the phenomena seems
to be common for the gapped theories \cite{zhit}. Secondly there is boojum negative energy
localized at the boundary which seems to be common phenomena as well. Moreover
it can be argued that two "wrong sign" phenomena are closely related. This relation
inspires the possible speculations on the relevance of the picture similar to
the topological insulators.
In that case one has the gapped bulk theory and the massless modes localized at the
boundary. It is not clear if there are massless boundary modes in our case as well
and this point certainly deserves the special attention. 

The third general feature concerns the phase transition as the function of the length
of the interval or the disc radius. This phase transition is the non-supersymmetric
counterpart of the marginal stability walls in  the twisted mass space in  the supersymmetric theories. 
 It that
case there is no phase transition but the spectrum
of the stable particles get changed at the wall. In the brane framework
the particles are represented by the strings or branes stretched between branes
and the position of the  marginal stability wall roughly  corresponds to
the particular distance
between the "background" branes in the internal space. It would be interesting
to relate the phase transition at some scale in the physical space
found in our case and the wall crossing
phenomena for the stretched strings at some scale in the internal space.

It would be interesting to discuss more general boundary conditions along the lines
developed in \cite{unsal}. We used the Dirichlet boundary condition  but
the nontrivial set of  open Wilson lines could be
included into consideration both for the strip and disc geometries. These
additional parameters could be very useful in investigation of the role of
non-commutativity of large volume and large N limits discussed in \cite{asorey}.
Another possibility to generalize the boundary condition is to include into
the game the boundary S-matrix.
Finally, it would be also interesting to investigate the effects discussed in this paper in the
framework  of cosmic strings reconnection  \cite{hashimoto}.

We thanks to M. Shifman, M. Voloshin and A.Yung for the useful comments.
A.G. thanks FTPI at University of Minnesota
where the part of the work has been carried out for  hospitality and support.
A.M. is partly supported by grant of Ministry of Education and Science of the Russian Federation under contract 8207, by
grant NSh-3349.2012.2.
The work  was supported in part by the grants
RFBR-12-02-00284 and PICS-12-02-91052.

\end{document}